# Big Data Architecture in Czech Republic Healthcare Service: Requirements, TPC-H Benchmarks and Vertica

Running head title:

## Vertica in Czech Republic Healthcare

Authors:

*Martin Štufi[a], Boris Bačić[b] and Leonid Stoimenov[c]*

[a]*Solutia, s.r.o., Prague, Czech Republic;*

[b]*School of Engineering, Computer and Mathematical Sciences, Auckland University of Technology, Auckland, New Zealand;*

[c]*Faculty of Electronic Engineering, Computer Science department, Niš, Serbia*

**Keywords:** Big data analytics (BDA), big data processing platform, NoSQL database cluster, Data Integration (DI), Master Data Management (MDM), Data Visualisation (DV).

**Corresponding author:** Martin Štufi, martin.stufi@solutia.cz

Abstract

Big data in healthcare has made a positive difference in advancing analytical capabilities and lowering the costs of medical care. In addition to providing analytical capabilities on platforms supporting current and near-future AI with machine-learning and data-mining algorithms, there is also a need for ethical considerations mandating new ways to preserve privacy, all of which are preconditioned by the growing body of regulations and expectations. The purpose of this study is to improve existing clinical care by implementing a big data platform for the Czech Republic National Health Service. Based on the achieved performance and its compliance with mandatory guidelines, the reported big-data platform was selected as the winning solution from the Czech Republic national tender (Tender Id. VZ0036628, No. Z2017-035520). The platform, based on analytical Vertica NoSQL database for massive data processing, complies with the TPC-H[1] for decision support benchmark, the European Union (EU) and the Czech Republic requirements, well-exceeding defined system performance thresholds. The reported artefacts and concepts are transferrable to healthcare systems in other countries and are intended to provide personalised autonomous assessment from big data in a cost-effective, scalable and high-performance manner. The implemented platform allows: (1) scalability; (2) further implementations of newly-developed machine learning algorithms for classification and predictive analytics; (3) security improvements related to Electronic Health Records (EHR) by using automated functions for data encryption and decryption; and (4) the use of big data to allow strategic planning in healthcare.

Introduction

Big data has influenced the ways we collect, manage, analyse, visualise and utilise data. For healthcare, adopting an eSystem with implemented big data analytics (BDA), there is an expectation that modern, robust, high-performance and cost-effective BDA technologies can preserve patient privacy while enhancing data-driven support for medical staff, as well as the broader patient population. Currently, the Czech Republic is in the process of adopting an eSystem in healthcare, leveraging BDA to enhance the quality of care with integrated national and regional support.

The scope of this paper is to report on the prerequisite factors influencing the implementation of big data architecture with the performance required to support the national strategy for the BDA adoption in the healthcare system of the Czech Republic. The reported healthcare solution, had to pass more than 100 complex requirements, pre-requisites and system conditions which were tested on the proposed platform, in compliance with European Union and Czech Republic regulations as well as the Transaction Processing Performance Council's TPC-H benchmarks.[1] In the authors' view, which aligns with global trends and EU initiatives:[2-6] (1) the growing amount of data in the healthcare, as well as data streaming IoT devices and mobile apps have made the adoption of BDA technologies inevitable for modern society; (2) combining BDA and AI with healthcare applications is a crucial step in advancing towards the next generation of healthcare eSystems; and (3) regarding replicability and knowledge transfer, the BDA platform implementation in one of the EU member states will shape decisions for other EU members who are in the process of transforming their healthcare systems.[2, 7]

## Multidisciplinary Background

Big data technologies have been adopted in many industries such as transport, banking, automotive, insurance, media, education and healthcare.[7] Common to the exponential trend of Internet network traffic, the volume of data produced every day is also increasing exponentially in modern healthcare. When the volume of data grows beyond a certain limit, traditional systems and methodologies can no longer cope with data processing demands or transform data into a format for the task required. Traditionally, small data portions as parts of online transaction processing (OLTP) systems are collected in a controlled manner, known as short atomic transactions.[8] In contrast, in big data clustered environments, there are stream and batch data processing, all requiring more flexibility for various data distribution patterns and matching eSystems scalability.

Typically, for big-data eSystems, stream-processing is concerned with (near) real-time analytics and data prediction, while batch data processing deals with implementing complex business logic with advanced and specialised algorithms.

Small data systems typically scale vertically by adding more resources to the same machine; this can be costly and eventually reach maximum possible upgrades. Contrastingly, big data systems are cluster-based and therefore depend mostly on horizontally scalable architecture, which in the long run provides increased performance efficiency at a lower cost by employing commodity hardware.

## Big Data Technology Perspective

The idea of applying big data cluster to process and analyse healthcare data is not new.[9-12] For example, in 2009, early experiments conducted on a 100-nodes cluster with a set of benchmarks, revealed various trade-offs in performance for selected parallel systems for storing and processing data intended for healthcare use.[13]

Recently, there has been a growing interest and need for eSystem platforms and cloud-based technologies emphasising new and innovative big data tools employing various data mining, machine learning[14, 15] and other AI-based techniques that could enable knowledge discovery, personalised patient-cantered modelling, identification of groups sharing similar characteristics, predictive analytics, improved drug safety and providing enhanced diagnostic capabilities.

Challenges and Opportunities

The integration of big data technologies in healthcare has local and global implications in terms of challenges and opportunities. Challenges in healthcare include "issues of data structure, security, data standardisation, storage and transfers, and managerial skills such as data governance".[16]

The use of AI, machine learning, personalised (targeted learning) models, predictive models and variable rankings can assist precision medicine and treatment risk/benefits assessments.[15] AI has the potential to automate the technical requirement for results from data to be *traceable* (inferring expert involvement, p.236 and using *prescriptive analytics* to infer explanations coupled with suggestions on an optimal set of actions, p.241).[17] As in the near future AI is expected to explain generated diagnostic assessments from data, big data technology can be both a catalyst and enabler for research advancements in the field. For example, an AI-based prototype capable of providing explainable diagnostic assessment coupled with traceable suggested actions (communicated as coaching cues) was implemented using captured kinematic data from tennis activity.[18] The personalised tennis coaching (PTC) system also includes different criteria for personalised and group (or skill-level) diagnostic assessments embedded in machine learning models.[18, 19] In a similar study, a neural network was unable to explain how it identified 'good' and 'bad' swings; however, it was able to demonstrate a flexible diagnostic assessment for novices and intermediate levels on the same observed motion dataset.[20] To design the PTC system to be able to generate comprehensible explanations, its (machine-learning) problem space was matched with human expertise-driven subspace modelling (for assessing adherence to descriptive coaching rules).[18]

To draw a parallel across healthcare, coaching and rehabilitation, let us consider on- and off-line data sources. Big data technologies can extract information and provide real-time analytics from healthcare records as well as processing large data streams that are pertinent to health, activity, and general wellbeing, all of which can be generated from a growing number of sensors found in mobile apps, medical IoT and wearable devices. If we consider data streams associated with body function (sleep, heartbeat, movement coupled with the ambient/environment and location) in our multidisciplinary approaches, produced research outcomes are likely to be transferable and strongly linked to the healthcare context.

Therefore, it is possible to connect healthcare and human motion modelling and analysis (HMMA) research into areas such as:[21] rehabilitation, sport, active life,[22] urban planning,[17, 22, 23] and ageing gracefully (including improvements in elderly care facilities).

Common to big data and the above multidisciplinary contexts are open questions regarding privacy and data protection pertinent to healthcare and ethical considerations.[2, 5, 24] In addition to privacy, safety and voluntary consent, which are common to address in academic data collection ethics, there are a number of issues transferrable to the contexts of big data and healthcare. Examples include[25-27]: legal aspects of data ownership, access, sensitive information, potential exploitation and data misuse, data collections by employer, insurance companies, gambling parties, and other stakeholders combined with wearable technology in sport analytics,[28] preventative medical benefits, optimisation of human performance while reducing the risk of injury. Thus, if legislation is lagging behind technological advancements and current trends towards private-public sector partnerships[2, 3] it is important to consider possible opportunities for exchange of sensitive information.

For a start, we may consider implementation of safety protocols and the exchange of anonymised metadata or other sensitive information to be restricted within the national legal boundaries. Anonymised data, such as video filtering while preserving diagnostic information, is of interest for online coaching and elderly care facilities. If we consider an example in which a patient falls overnight in the bathroom, unable to call for help, having a video stream via IoT device of only the silhouette or stick figure rendition of the patient would allow an AI-based medical alert to be triggered autonomously while preserving the privacy of the vulnerable user population. A computer vision systems that can extract a pseudo 2D and 3D silhouette have been developed and evaluated for diagnostic implications in augmented video golf coaching.[29, 30] There is also a growing body of literature in deep learning and openCV (https://opencv.org) with open-source software available to enable video filtering and human pose estimation as stick figure, most of which are capable of running on low-cost hardware or embedded Linux IoT devices. Although there are not many research papers regarding the use of open source software for augmented video coaching and rehabilitation,[31-33] it is important to know that there are open-source software tools that would not require users to 'share' their videos on cloud, or frequently 'nudge' end-users to upgrade their apps in order to manage their videos or perform common tasks such as streaming/previewing, saving or extracting a video sequence of interest.

To enable improvements to the quality and delivery of healthcare services, the implementation of big data architectures combined with data analytics to available EHR[34], there is potential to:

- improve the quality of personalised care and medical services;
- reduce cost of treatment;
- use predictive analytics for e.g. patients' daily (loss of) income and disease progression;
- use real-time visualisation and analytics for immediate care and the cases of readmission;[35]
- enhance patient engagement with their own healthcare provider via processing satisfaction evaluation data and self-reported health status;[6, 36, 37]
- reduce the occurrence of fraud and identify high-cost profiles that may require more healthcare resources than most of the population;[6, 38] and
- use healthcare data for identification of trends, strategic planning, governance, improved decision-making and cost reduction.[16, 35]

To enable advancements towards the next generation of BDA platform that can help and improve healthcare outcomes, this study addresses the following questions:

1. It is possible to design and build a BDA platform for the Czech Republic healthcare service, in line with EU legislation, TPC-H[1] benchmarks and other statutory requirements?
2. If so, what BDA platform would provide optimal cost and performance features while allowing installation of open-source software with various machine learning algorithms, development environments and commercial visualisation and analytical tools?
3. To what extend would such a BDA-based eSystem be future-proof for maintaining reliability, robustness, cost-effectiveness and performance?

## Industry Benchmarks

Industry benchmarks have an important role in advancing design and engineering solutions in database systems. For example, the Transaction Processing Performance Council (TPC),[1] has an important role in encouraging the adoption of industry benchmarks in computing, which are today widely used by many leading vendors to demonstrate their products' performance. Similarly, large buyers often use TPC benchmark results[39, 40] as a measurable point of comparison between new computing systems and technologies to ensure a high level of performance in their computing environments.[41]

## Big Data Analytics

Analytical technologies for big data[23] are showing promising results in their attempts to manage ever-expanding data in healthcare. For example, a 2014 Massachusetts Institute of Technology (MIT) study on big data in intensive care units[12] reported findings that data analysis could positively predict critical information, such as duration of hospitalisation, number of patients requiring surgical intervention and which patients could be at risk of sepsis or iatrogenic diseases. For such patients, data analytics could save lives or prevent other complications that patients might encounter.

Technologies utilising BDA are also being successfully employed outside of hospitals.[42] The medical community and government bodies now recognise the importance of monitoring the incidence of influenza illness using massive data analysis technologies.[43] Seasonal influenza epidemics are a significant problem for public health systems, annually leading to 250,000-500,000 deaths worldwide.[44-46] Furthermore, new types viruses against which population lacks immunity can lead to a pandemic with millions of deaths.[44] Early detection of disease activity leads to a faster response and may reduce the impact of both seasonal and pandemic influenza in terms of saving lives or reducing respiratory illnesses on world-wide scale.[44] One method of early detection is to monitor Internet search behaviour in relation to health queries such as employed by Google.[14, 46] In addition, it was discovered that some queries are strongly correlated with the percentage of doctor visits when the patient presents symptoms of influenza. This correlation made it possible for Google to produce an algorithm which estimates influenza activity in different regions of the United States with a one-day delay. Among other algorithms, this approach allows Google to use queries to detect epidemics from influenza-like searches in areas where population has regular access to the Internet.

### SQL vs NoSQL Approaches

Data can be stored and processed in either a row-oriented or column-oriented format. The row-oriented principle based on Codd's relational model is well-established in most database applications.[47-50] However, such well-established relational database management systems (RDBMS)[49, 51] are not efficient for analytical applications that mostly perform create, read, update and delete operations. Over the last years, NoSQL[40] databases have been tested and studied as well as their performance has been evaluated in many different studies,[52, 53] where some of the studies focused their evaluations on the advantages of the use of NoSQL technologies.[54] For BDA platforms architects, known differences between Structured Query Language (SQL) and NoSQL database management systems makes a design a challenging task with a number of decisions to address the purpose and related set of requirements. Newer than SQL, NoSQL databases support the notion of *elastic scaling*, allowing for new nodes to be added to improve availability, scalability, and fault tolerance.[55]

Many of related work and reviews for big data techniques[56] and technologies used in healthcare relay mostly on silo principle for data integration, data processing, and data visualisation. An application which operates on the columns in the dataset allows overcoming performance problem with "NoSQL" or "Not Only SQL" database.[44, 46] In nowadays, these databases we can recognise on premise or in the cloud. Cloud computing[48, 57] offers this database service too. NoSQL databases provide elastic and horizontal scaling features allowing new nodes to be added. New nodes are typically designed based on low-cost (*commodity*) hardware.

In relation to the main objective of this work, produced real-life platform for big data integration, master data management, ad-hoc analysis, data storing, data processing, and visualisation is based on NoSQL database for data storing and data processing on Vertica clusters.

## Materials and Methods

The dataset used for this cross-sectional study includes an anonymised real-life data sample provided by the Czech Republic healthcare institute (IHIS). To comply with the two-phase IHIS acceptance testing, the requirement analysis was combined with the design science approach[52] involving the production of the scalable platform, architecture, software and hardware infrastructure. The produced solution, based on the cyclic experimental design approach, has met all the requirements while also achieving the highest performance ranking.

Incremental performance and functionality improvements from phase I and phase I+II evaluations involved:

1. Phase I-compliant big data eSystem requirements and design decisions influencing architecture design;
2. Decomposition of acceptance testing requirements;
3. System optimisation to of a set of the requirements (N=112) for weighted scoring, including TPC-H decision support, and minimal system performance; and
4. Performance-driven eSystem optimisation from available test datasets.

The performance evaluation is based on database management systems[45] capable of tackling big data such as Cassandra, CouchDB, HBase, Impala,[58] MongoDB[49] and Vertica.[59]

IHIS Requirements

The IHIS requirements can be grouped by the following aspects:

- **Scalability:** the eSystem must allow performance enhancement via additional and accessible computing technology, including commodity hardware products.

- **Modularity, Openness, and Interoperability:** the system components must be integrated via specified interfaces according to exact requirement specifications. It is also essential that a wide variety of vendors can readily utilise system components.

- **Exchangeability:** the eSystem solution must support installation of open-source operating systems and contain tools for non-profit and educational purposes. The eSystem must comply with standard Data Warehouse (DWH) systems. Some components must be interchangeable with Massive High Data Analytics (MHDA) system components.

- **Extensibility:** all tools and components of the eSystem must provide space for future upgrades, including functionality and capability advancements.

- **Quality Assurance:** a tool for validating data and metadata integrity is required to ensure that processed data remains accurate throughout the analysis procedure.

- **Security:** the eSystem must be operable on local servers, without reliance on cloud or outsourced backup systems. It is essential that the eSystem provides security for all data against external or internal threats. Therefore, authorisation, storage access and communication are of utmost concern. User access rights had to be set to the database, table or column level to restrict data access to a limited number of advanced users. The eSystem must log all executions and read operations for future audits. The eSystem must support tools for version control and development, while meeting the requirements for metadata and data versioning, backup and archiving.

- **Simplicity:** the eSystem must allow for parallel team collaboration on all processes, data flow and database schemas. All tasks must be fully editable, allowing *commit* and revert changes in data and metadata. It is essential that the eSystem be simple and easy to use, as well as stable and resilient to subsystem outages.

- **Performance:** the eSystem must be designed for the specified minimum number of concurrent users. Batch processing of data sources and sophisticated data mining analyses are considered essential. Complete data integration processing of quarterly data increments must not exceed one hour.

The most important IHIS requirements mandate that:

1. All tools, licenses, and environmental features used in the Proof of Concept (PoC) tests must match the eSystem offer submitted and documented in the public contract. To meet contractual obligations, the proposed solution cannot have: insincerely increased system performance, altered available license terms or otherwise improved or modified results vis-à-vis the delivery of the final solution. The environment configuration must satisfy the general requirements of the proposed eSystem (usage types, input data size, processing speed requirements).

2. The proposed eSystem cannot be explicitly (manually) optimised for specific queries and individual task steps within a test. The test queries are not to be based on general metadata (cache, partitioning, supplemental indexes, derived tables and views), except in exceptional cases where optimising the loading of large amounts of data is needed. The techniques based on general metadata may be used in future for enhancing performance but are not required as a precondition for system availability. To load large amounts of data, the environment configuration can be manually adjusted to a non-standard configuration for further test steps (Fig. 1).

3. The configuration must not be manually changed during the test to optimise individual tasks - the eSystem is required to be universal for tasks that may overlap in time.

TPC-H Performance Requirements and Tests

Meeting TPC-H benchmarks involves testing for minimum requirements including a set of values and parameters. Standardised test conditions specified in the TPC-H Benchmark™ are available online (http://www.tpc.org/tpch). The IHIS requires that any proposed eSystem meets performance metrics aligned with standard TPC-H workloads during developmental phase testing (Fig. 1).

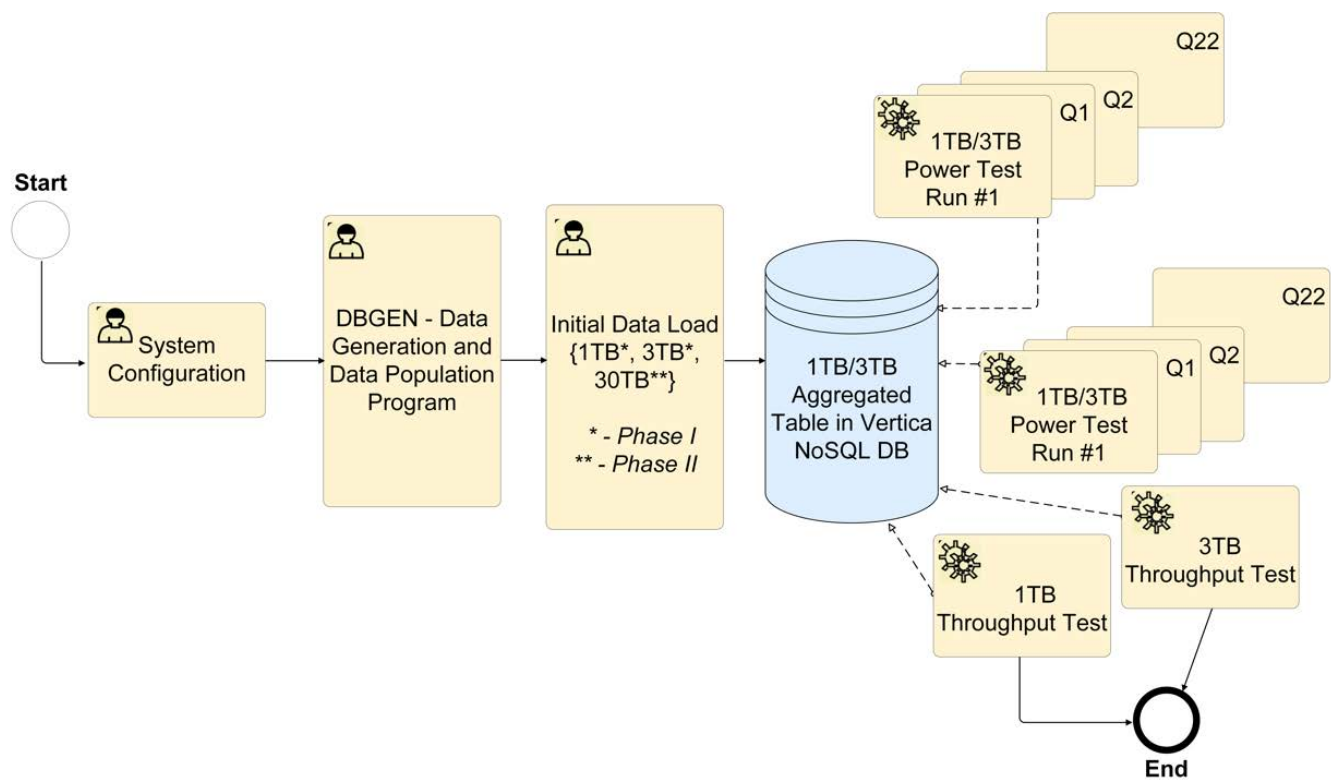

Fig. 1. Testing activity diagram for the data warehouse eSystem configuration.

For data storage benchmarks, the data must be stored on independent disks, with a replication factor more significant than two. The solution must also support best practices regarding data security and data protection, including hot backup, cold backup and recovery.

Table 1 shows predefined TPC-H parameters by IHIS for initial test datasets (1TB and 3TB), and values for the *Power Tests* in $1^{st}$ and $2^{nd}$ run for 1TB and 3TB limits. Before starting and testing TPC-H benchmarks, optimising the system for specific queries (such as manual or another non-standard optimisation) is not permitted.

Table 1. TPC-H parameters predefined by IHIS for initial imports for data size (1TB and 3TB), including limiting values for Power Tests in 1st and 2nd runs for 1TB and 3TB

| Parameter | Limit [hours] | Achieved results [hours] |
|---|---|---|
| Initial import TPC-H 1 TB | 24 | 2.94* |
| Initial import TPC-H 3 TB | 96 | 5.99* |
| Power test TPC-H 1TB – 1st run | 1.5 | 1.4** |
| Power test TPC-H 1TB – 2nd run | 1.5 | 1.36** |
| Power test TPC-H 3TB – 1st run | 5 | 4.2** |
| Power test TPC-H 3TB – 2nd run | 5 | 4.17** |

Note:
* - Initial import of 1TB and 3TB data results (Table 2);
** - TPC-H benchmark (Table 3).

The TPC-H tests are to emulate the future production eSystem behaviour. For all contenders (Tender Id. VZ0036628, No. Z2017-035520), test data consisted of simulated medical documentation records from three fictional insurance companies: three standard quarterly packages (one quarter for each company), plus one correction (simulating a situation where one insurer supplied inadequate data). Data batches (compressed using the ZIP format) were exchanged in real time containing images and structured alphanumerical data in the comma-separated (CSV) format. Standard input data were up to 30GB per packet, amounting to a total 3 TB of data. Test data contained roughly the same number of rows of expected data, but with a reduced number of columns and with added redundant attributes to reflect the problem dimensionality and approximate amount of data anticipated. Data related to patient drug use was confirmed with *The Anatomical Therapeutic Chemical Classification System.*[60]

For the purpose of conducting TPC-H Benchmark testing, a contender's MHDA system had to be installed *on premise* utilising a private network. The on-premise multi-user MHDA system must operate in a parallelised application environment. Once the metadata are loaded, the system was required to run without intervention to prevent any configurations being altered manually, thus compromising the TPC-H tests integrity.

The installed operating system chosen by Solutia was Red Hat Enterprise Linux Server release 6.8 (Santiago). Our solution met all of the requirements for Massively Parallel Processing (MPP) as an MHDA system. The proposed architecture also allowed for both remote supports according to the specified Service Level Agreement (SLA) for fault correction, and by the end of Next Business Day (NBD) requirements.

For step one of testing activities (Fig. 1), we configured the system architecture with five nodes operating in the Vertica 9.0.1-1 database cluster. For step two, a *DBGEN* program generated 1TB or 3TB databases. At this point, the initial data loaded into the system and we ran 1TB and 3TB *power* and *throughput* tests. These tests resulted in records of individual measurement results. After testing one cluster, we deleted the data and generated another 3TB database. We repeated these power tests for each of the five nodes, including the recording of measurement results.

To compute query processing power[1] for a database of a given size (TPC-H_$Power_{@Size}$), we used equation (1), in compliance with the most recent TPC Benchmark$^{TM}$ H standard specification (revision 2.18.0, p. 99)[1]:

$$\text{TPC-H\_}Power_{@Size} = 3600 * e^{\left\{-\frac{1}{24}\left[\sum_{i=1}^{22} \ln(QI(i,0)) + \sum_{j=1}^{2} \ln(RI(j,0))\right]\right\}} * SF \qquad (1)$$

where *QI*(*i*,0) is the timing interval, in seconds, of a query $Q_i$ within the single query stream of the TPC-H power test; *RI*(*j*,0) is the timing interval in seconds, of a refresh function *RFj* within the single query stream of the power test; and *SF* represents the corresponding scale factor of the database size.[1]

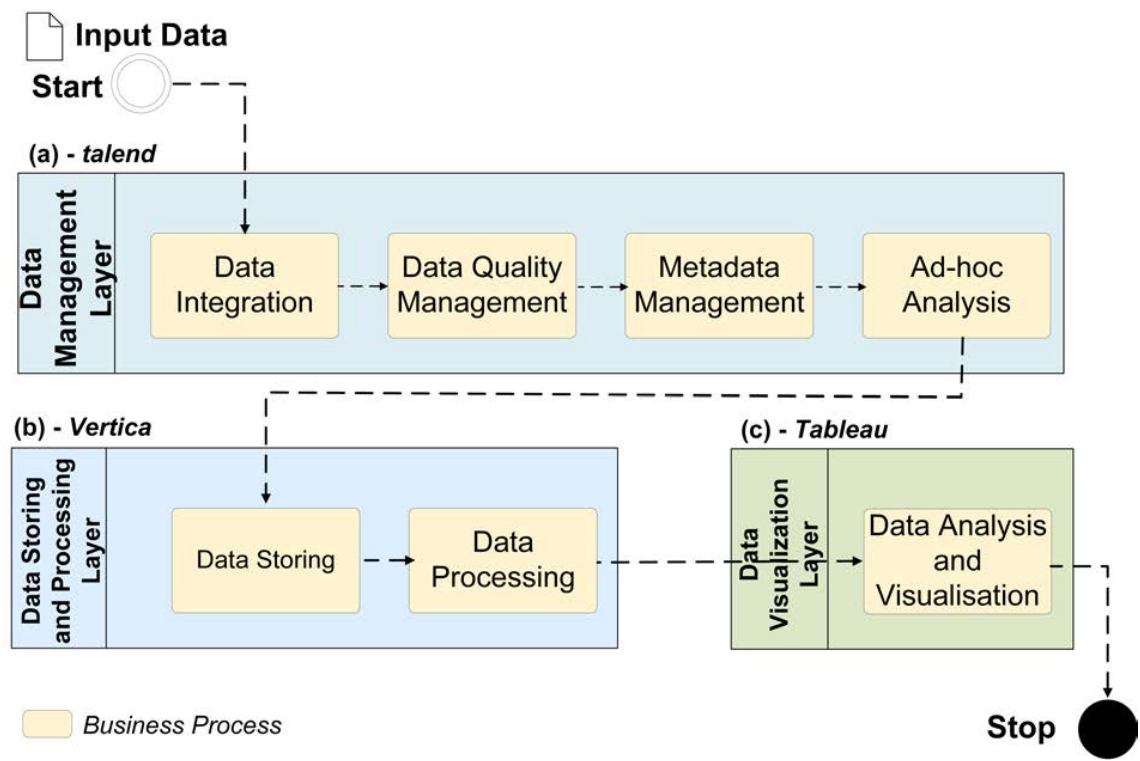

Fig. 2. Business process organisation and tests data flow.

Business processes organisation and data flow (Fig. 2) shows integration of Talend, Vertica and Tableau. Processing flow of the supplied test dataset starts with *Data Management Layer* (through the Data Integration, Data Quality Management, and Ad-hoc Analysis subparts), linked to *Data Storing and Processing Layer*. The final stage is the *Data Visualisation Layer* (for Data Visualisation and Analysis of the pre-processed data).

Vertica-Based eSystem Solution Architecture

Within the presented BDA platform, we distinguish the following logical sub-components of the system: Data Integration Layer (DI), Data Storage (DS), Ad-hoc Analysis Preparation (AAP), Data Quality Management (DQM) and Meta Data Management (MDM). The entire eSystem solution based on Vertica is designed to support the processing Massively Parallel Processing (MPP) database requirements.[57, 59, 61] Since the processing of big data requires high-performance computing, we used a cluster computing architecture to take advantage of massive parallel and NoSQL database.[55]

The Vertica Analytic Database enables the principle of C-Store project,[51] which is widely used as commercial relational database systems for business-critical systems. Vertica database has characteristics that are important for exceeding expected system performance while meeting all IHIS requirements such as: (1) massively parallel processing (MPP) system, (2) columnar storage, (3) advanced compression, (4) expanded cloud integration and (5) specialised tool for database design and administration. It is also important to note that Vertica effectively utilised built-in functionalities for an *analytic workload* (e.g. few to ten per second) rather than for a *transactional workload* (e.g. few hundreds to thousands per second).

To choose Vertica for our client's requirements, we also considered the following benefits:

- Vertica provides an SQL layer as well as it supports connection to Hadoop and fast data access to ORC, Parquet, Avro, JSON as column oriented data;
- Vertica affords very high data compression ration with high degrees of concurrency and massive parallelism for processing tasks;
- Vertica expanded analytical database support for Kafka, Spark;
- we considered Vertica's pricing model a good fit for enterprise solutions;
- Vertica is specifically designed for huge analytical workloads;
- Vertica allows cloud integration for the future development; and
- Vertica provides advanced compression capabilities which can handle and deliver high speed results data at petabyte scale.

Key Components Overview

The presented BDA platform, as a distributed and large-scale system, is designed on commodity hardware with gigabit Ethernet interconnections (Fig. 3). Adding nodes the Vertica database allows system performance improvement as per IHIS requirements and general expectations for exponential growth in healthcare data.

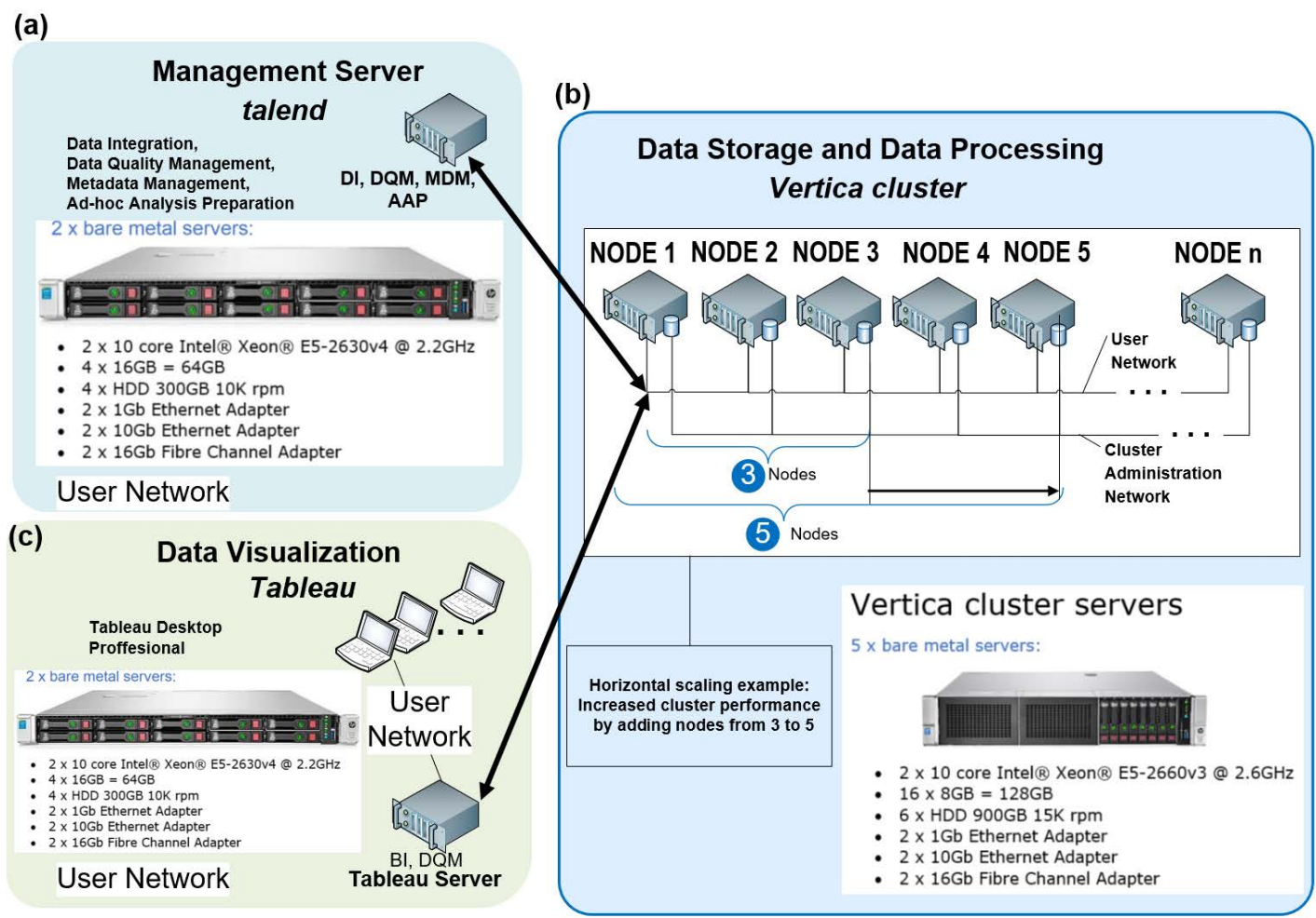

Fig. 3. Architecture and infrastructure diagram showing the key components of BDA platform.

The BDA platform unifies three key components Talend, Vertica and Tableau. As a specialised data integration environment for BDA platforms, Talend provides functionality for: Ad-hoc Analysis Preparation, Metadata Management, Data Quality Management, and Data Integration.

Vertica NoSQL database built on five nodes provides Data Storage and Data Processing. The Data Visualisation is covered by the Tableau Desktop (Professional edition).

### Data Integration (DI) Layer

The data integration (DI) layer represents a system module which enables parameterised data manipulation functions, including: data transformation, processing control and hierarchy, reading, writing and parallel or sequential tasks/threads processing. We use the term *metadata* to describe the resulting statistics, classification or data aggregation tasks. The DI provides metadata for development, test and production environments. The DI layer also provides visualisation of its processes in the form of data-flow diagrams. Another DI-specific tool generates outputs from pre-processed data. This tool also supports rapid process development including selection and transformation of large volumes of primary data in parallel multi-threaded execution. To deal with near-future technical and operational challenges, the DI module also contains a debugging tool for software development, testing and maintenance.

### Data Storage (DS)

The Data Storage (DS) represents a system module which contains cluster-based, horizontally scalable physical architectures built onto NoSQL Vertica databases. The DS runs on commodity hardware with distributed storage capabilities, which allows for Massively Parallel Processing (MPP) over the entire data collection. The DS keeps data in a column format in two containers, Write Optimised Store (WOS) and Read Optimised Store (ROS), for best performance. Each cluster is a collection of hosts (nodes) with Vertica software packages. Each node is configured to run a Vertica NoSQL database as a member of a specific database cluster, supporting: redundancy, high availability, and horizontal scalability, ensuring efficient and continuous performance. This infrastructure allows for recovery from any potential node failure by allowing other nodes to take control. For the presented solution (Fig. 3), we set a fault tolerance K-safety=2.[61] The DI components specify how many copies of stored data Vertica should create at any given time.

Data Quality Management (DQM)

The data quality management (DQM) module supports data quality control including trends and data structures. The DQM generates complex models for end-users supporting data analysis for error detection and correction as well as sophisticated visualisation and reporting required for quality control tasks. It creates, sorts, groups and searches for validation rules entered in a structured form. Validation rules can be executed over a user-defined dataset and managed centrally.

Metadata Management (MDM)

The metadata management (MDM) module supports the management of user, technical and operational metadata. The MDM centrally processes metadata from every component of the MDHA system, housed collectively in the data warehouse.

The MDM can compare different versions of metadata and display outputs including visualisation intended for data reporting. The MDM is able to create dynamic, active charts and tables allowing multidimensional and interactive views. The MDM uses sandboxing for testing temporary inputs and outputs and can generate outputs in HTML, PDF and PPT formats. The MDM component utilises Online Analytical Processing (OLAP) operations over a multidimensional data model. Additionally, it contains a glossary of terms and concept links, to enable impact and lineage analysis.

**Ad-hoc Analysis Preparation (AAP)**

For ad-hoc analysis preparation (AAP) processes, we programmed into the Talend Open Studio integration tool two different versions. In the first version, the MHDA uses Extract Transform and Load (ETL) components of the integration tool. These components read data from data warehouse structures (dimensions and fact tables) into memory. Then, the filtering and aggregation components process the data into an output table. The second version uses Extract Load and Transform (ELT) components of the integration tool. Both ETL and ELT components are able to generate user-friendly, unmodified SQL Data Manipulation Language (DML) statement(s) in the background. The AAP module accelerates the processing time without having to load large amounts of metadata into the program memory.

Figure 4 shows forecasting on a historical test dataset supplied by IHIS, where we used the ARIMA[62] in-database approach to timeseries. This model can be created either directly in the NoSQL Vertica database, which supports predictive modelling, or in a separate statistical tool such as Tableau, which will take data from the database and return the created model (written in Predictive Model Markup Language (PMML) or another format the database supports). Including analytics algorithms into the database often leads to increased processing demands on BDA platforms.

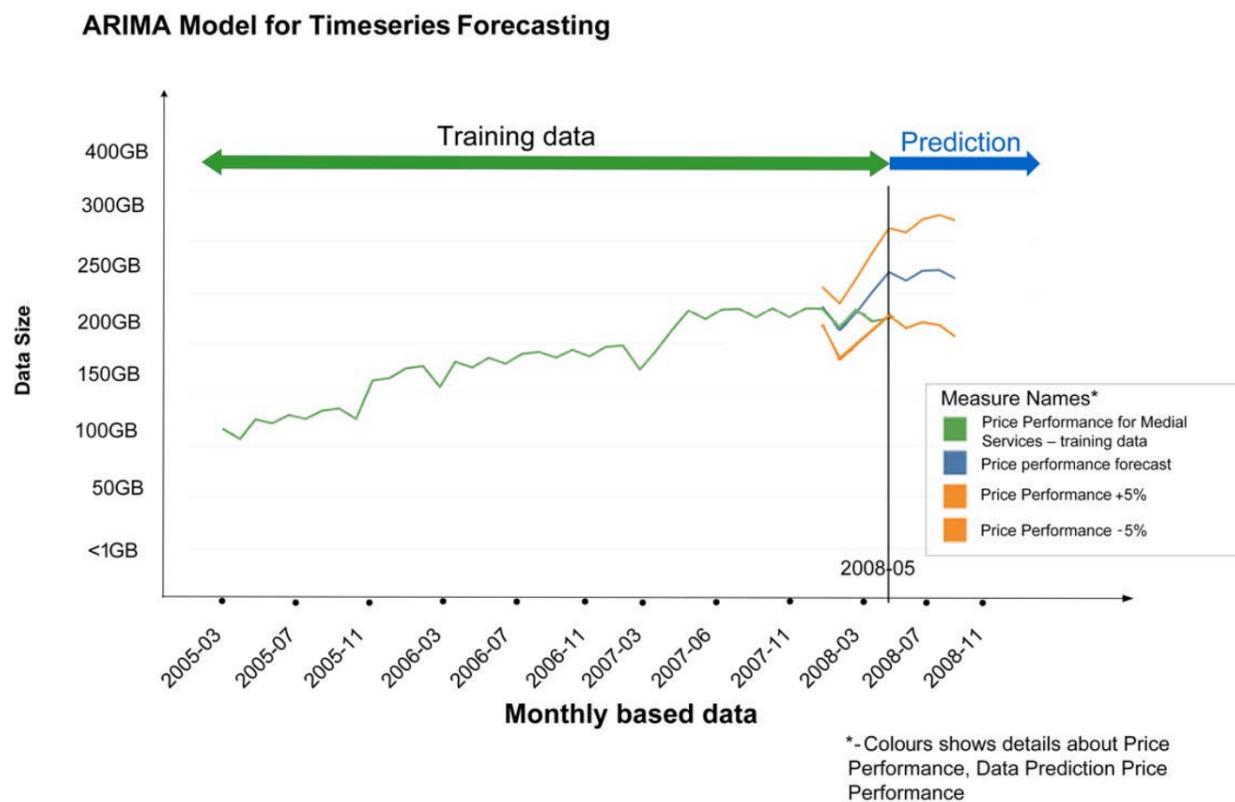

Fig. 4. Visualisation of predictive model for timeseries forecasting based on supplied test data.

Data Visualisation (DV)

The data visualisation (DV) module contains tools for describing data perspectives and knowledge discovery from data. The DV components represent data and metadata visually and give interpretations for possible insights. Additionally, we embedded DV components in Tableau to provide data and metadata visualisations in graphs and pictures. Tableau is a popular interactive analytical and data visualisation tool, which can help simplify raw data into easily comprehensible dashboards and worksheets. For example, Fig. 5, depicts a part of the data visualisations from one of the IHIS case studies with a geographical map overlay .

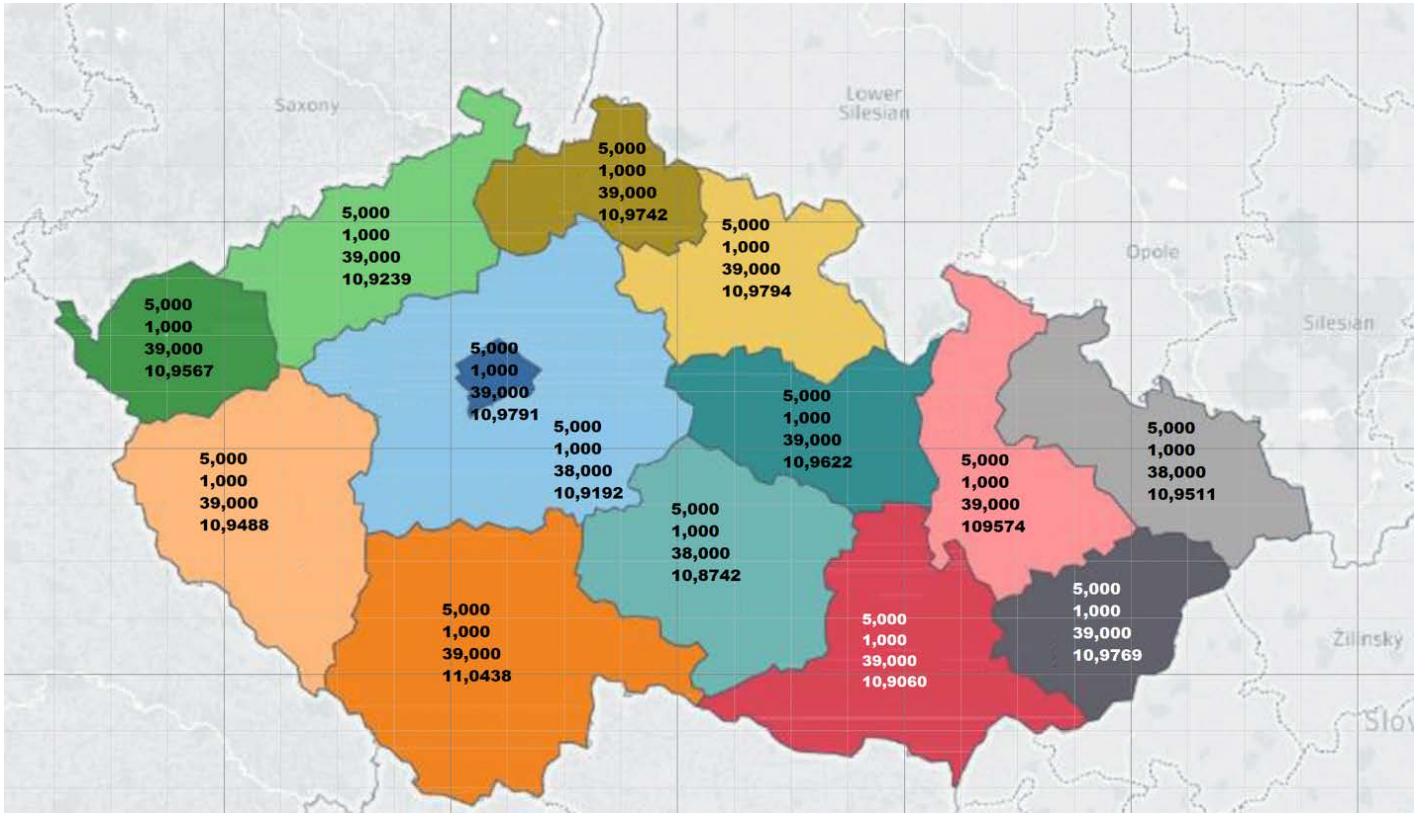

Fig. 5. Real-life diagnosis as regional data visualisation in the Czech Republic using Tableau Desktop.

We found that the DV combined with Tableau provide a powerful, secure and end-to-end analytics platform for data visualisation tasks.

Results

The presented winning solution as a proof of concept (PoC), was implemented and transferred to our client IHIS, which is integrated with the Ministry of Social and Labour Security, Ministry of Defence, Ministry of Internal Affairs, Ministry of Health Insurance and Eurostat (Statistical Office of the European Union). IHIS requirements also complied with the EU-based General Data Protection Regulation (GDPR).

## TPC-H Tests Configuration

TPC-H requires data to be generated for eight tables using the specified scale factor (*SF*), which determines the approximate amount of data in gigabytes (Fig. 6). We used the TPC-H power test which measures the throughput/response times of a sequence of 22 queries (defined on p. 29).[1] Vertica supports the ANSI SQL-99 standard and all queries are applied with no syntax changes. The test datasets were created by TPC-H DBGEN program (Fig. 1). In our tests, we found that the queries Q 9 and Q 21 are more complex in comparison with the commonly expected queries. For power benchmark purposes, we have shared TPCH_SF1000, consisting of the row size x1000 (several billion elements).

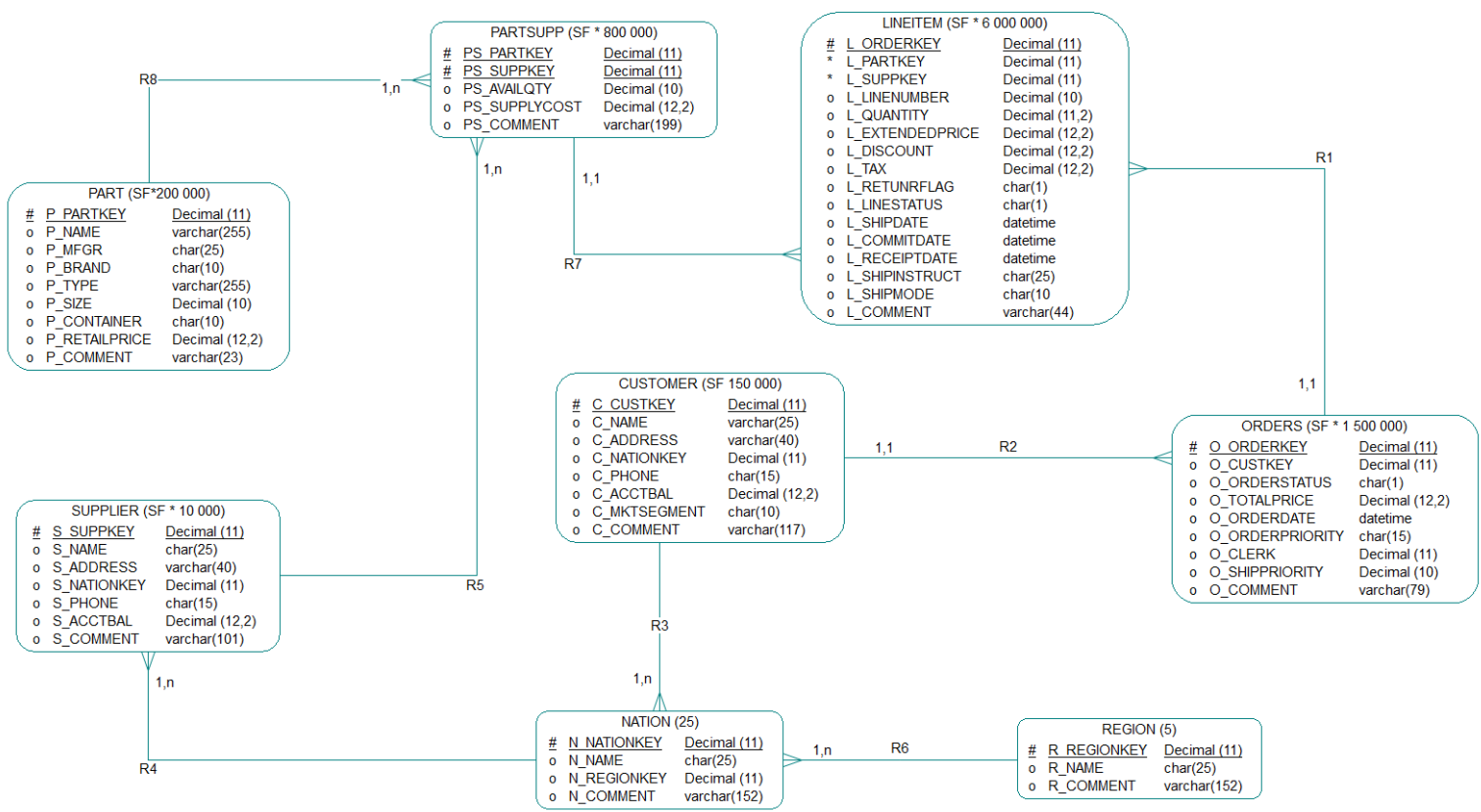

Fig. 6. The components of TPC-H consisting of eight tables (adapted from p.13).[1]

The performance achieved using the dataset predefined by TPC-H (Fig. 1) shows that the developed eSystem (as PoC) outperformed other competitors with similar product characteristics.[63-65] Experiments using the presented BDA architecture (Fig. 3) and reported performance (Table 2, Table 3, Fig. 7, and Fig. 8) were also tested by the government. The developed eSystem was installed within the Czech Republic borders in *on-premise* centralised mode using data communication channels that are physically separated from the existing Internet infrastructure.

Table 2. Measured results compliant with TPC-H benchmarks for initial import of 1TB and 3TB data

| Data Size | 1TB data | | | 3 TB data | | |
|---|---|---|---|---|---|---|
| Table | No. of rows | Duration in [s] | Duration in hours | No. of rows | Duration in [s] | Duration in hours |
| **Customer** | 150,000,000 | 1,185.00 | 0.33 | 450,000,000 | 4,100.00 | 1.14 |
| **Nation** | 25 | 0.10 | 0.00 | 25 | 0.20 | 0.00 |
| **Orders** | 1,500,000,000 | 2,533.00 | 0.70 | 4,500,000,000 | 5,423.00 | 1.51 |
| **Part** | 200,000,000 | 272.00 | 0.08 | 600,000,000 | 865.00 | 0.24 |
| **Part supp.** | 800,000,000 | 1,342.00 | 0.37 | 2,400,000,000 | 4,240.00 | 1.18 |
| **Region** | 5 | 0.07 | 0.00 | 5 | 0.07 | 0.00 |
| **Supplier** | 10,000,000 | 105.00 | 0.03 | 30,000,000 | 266.00 | 0.07 |
| **Line item** | 5,999,989,709 | 10,594.00 | 2.94 | 18,000,048,306 | 21,548.00 | 5.99 |
| **Total load duration in hours** | 2.94[*][h] (16,031.17[s]) | | | 5.99[**][h] (36,442.27[s]) | | |

\* - Measured results for 1TB dataset generated by DBGEN and imported into NoSQL Vertica DB;
\*\* - Measured results for 3TB dataset generated by DBGEN and imported into NoSQL Vertica DB.

Table 3. TPC-H benchmark queries for 1 TB and 3 TB test data

| Data size | 1 TB data | | | | 3 TB data | | | |
|---|---|---|---|---|---|---|---|---|
| Query No. | 3 nodes | 3 nodes | 5 nodes | 5 nodes | 3 nodes | 3 nodes | 5 nodes | 5 nodes |
| | 1st run | 2nd run | 1st run | 2nd run | 1st run | 2nd run | 1st run | 2nd run |
| Q1 | 51 | 267 | 232 | 161 | 427 | 383 | 441 | 457 |
| Q2 | 22 | 23 | 19 | 15 | 52 | 42 | 36 | 40 |
| Q3 | 65 | 64 | 55 | 40 | 128 | 109 | 121 | 125 |
| Q4 | 480 | 470 | 287 | 320 | 918 | 897 | 914 | 900 |
| Q5 | 114 | 177 | 71 | 70 | 484 | 462 | 465 | 454 |
| Q6 | 0.7 | 0.6 | 0.8 | 0.5 | 1.2 | 1 | 1.4 | 1 |
| Q7 | 129 | 119 | 65 | 65 | 144 | 129 | 140 | 140 |
| Q8 | 34 | 37 | 40 | 22 | 375 | 361 | 270 | 263 |
| Q9 | 2551 | 2576 | 1555 | 1397 | 16015 | 15173 | 3791 | 3824 |
| Q10 | 130 | 52 | 44 | 72 | 65 | 58 | 64 | 64 |
| Q11 | 7 | 6 | 5 | 3.8 | 13 | 10 | 10 | 11 |
| Q12 | 13 | 13 | 8 | 11 | 24 | 20 | 23 | 23 |
| Q13 | 237 | 221 | 180 | 136 | 296 | 251 | 325 | 301 |
| Q14 | 55 | 49 | 41 | 36 | 111 | 102 | 105 | 105 |
| Q15 | 7 | 7 | 4 | 4 | 9 | 7 | 9 | 10 |
| Q16 | 42 | 42 | 29 | 30 | 87 | 78 | 86 | 88 |
| Q17 | 12 | 11 | 8 | 6 | 23 | 19 | 23 | 22 |
| Q18 | 380 | 376 | 517 | 554 | 741 | 721 | 743 | 746 |
| Q19 | 58 | 58 | 41 | 41 | 112 | 104 | 111 | 110 |
| Q20 | 131 | 129 | 87 | 73 | 156 | 137 | 150 | 147 |
| Q21 | 1763 | 2850 | 1703 | 1787 | 7278 | 7021 | 7196 | 7085 |
| Q22 | 60 | 67 | 62 | 35 | 107 | 88 | 110 | 106 |
| Results in seconds | 6341.7 | 7614.6 | 5053.8 | 4879.3 | 27566.2 | 26173 | 15134.4 | 15022 |
| Results in hours | 1.76 | 2.12 | 1.4 | 1.36 | 7.66 | 7.3 | 4.2 | 4.17 |

Note (Q# - Query name):[1]
Q1 - Pricing Summary Report, Q2 - Minimum Cost Supplier, Q3 - Shipping Priority,
Q4 - Order Priority Checking, Q5 - Local Supplier Volume, Q6 - Forecasting Revenue Change, Q7 - Volume Shipping, Q8 - National Market Share, Q9 - Product Type Profit Measure, Q10 - Returned Item Reporting, Q11 - Important Stock Identification, Q12 - Shipping Modes and Order Priority, Q13 - Customer Distribution, Q14 - Promotion Effect, Q15 - Top Supplier, Q16 - Parts/Supplier Relationship, Q17 - Small-Quantity-Order Revenue, Q18 - Large Volume Customer, Q19 - Discounted Revenue, Q20 - Potential Part Promotion, Q21 - Suppliers Who Kept Orders Waiting, Q22 - Global Sales Opportunity.

The performance of TPC-H tests running on a Vertica cluster for 1 TB and 3 TB data size (Table 3) are visualised in Fig. 7 and Fig. 8, indicating similar duration patterns for complex and commonly expected queries.

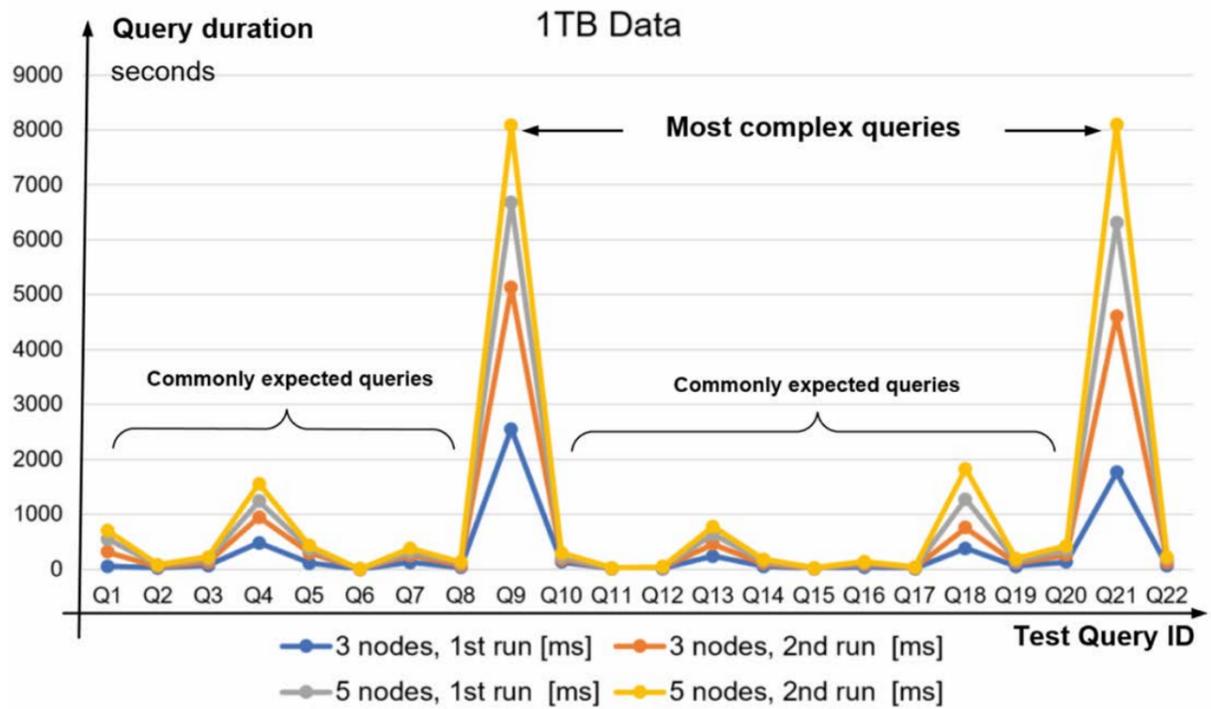

Fig. 7. TPC-H query duration on 1TB database (from Q1 to Q22).

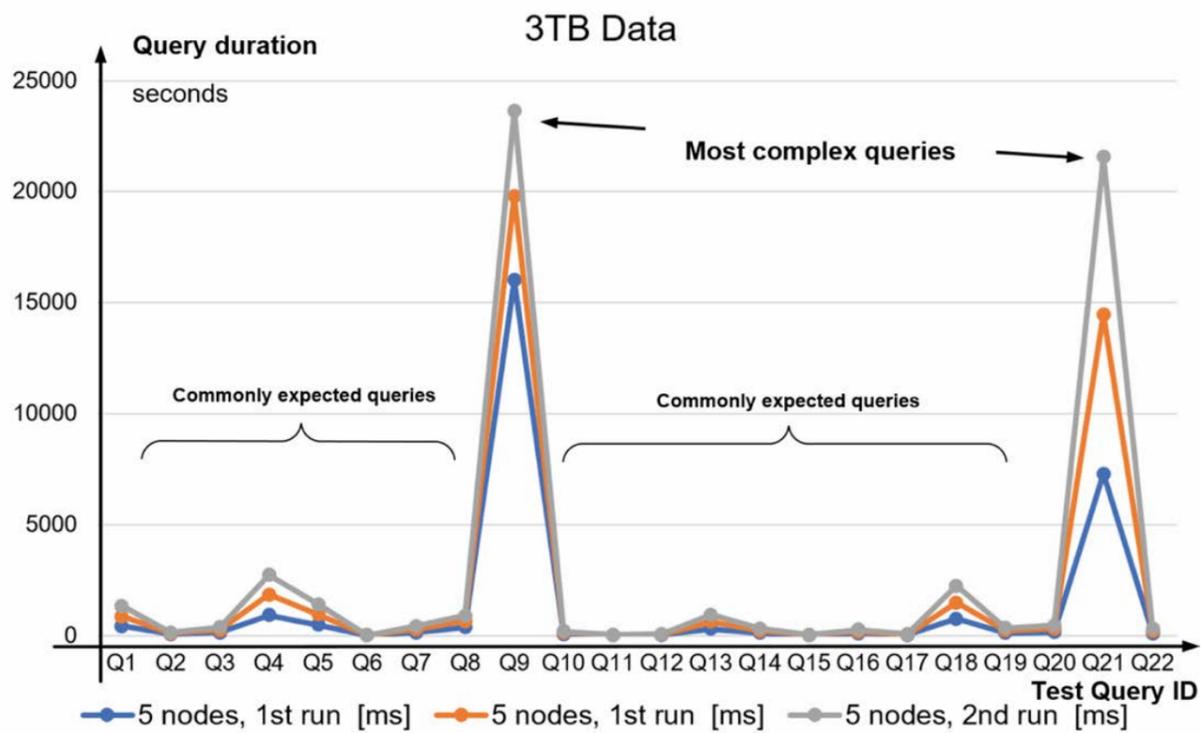

Fig. 8. TPC-H query duration on 3TB database (queries Q1 to Q22).

Monitoring I/O requests to accurately capture workload behaviour is important for the design, implementation and optimisation of storage subsystems. The TPC-H trace collection on which we conducted the analysis, was collected on Vertica 9.0.1 database cluster running on CentOS Linux 7.3, five nodes, 2x10 cores CPU Intel® Xeon E5-2660v3@2.66 GHz, 16x8 GB = 128 GB RAM, 6xHDD 900 GB(@15K rpm), 2x1 Gb Ethernet, 2x10 Gb Ethernet, 2x16 Gb Fibre Channel Adapter.

As introduced, the TPC-H can also be used as a metric to reflect on multiple aspects of a NoSQL Vertica database system's ability to process queries. The aspects of performance improvements for different database sizes and system expansion are captured collectively in Table 4, Fig. 9, Fig. 10 and Fig. 11. As such, it is possible to infer anticipated needs for future system upgrades and expected performance based on evidence from measured performance improvements from three to five nodes tested on 1 TB and 3 TB databases.

Table 4. Vertica cluster on three and five nodes in the 1st and 2nd run on 1TB and 3TB

| Query No. | 3 nodes 1st run | 3 nodes 2nd run | 5 nodes 1st run | 5 nodes 2nd run |
|---|---|---|---|---|
| Results in [s] for 1TB [s] | 6 341.7 | 7 614.6 | 5 053.8 | 4 879.3 |
| Results in [s] for 3TB [s] | 27 566.2 | 33 099.26 | 15 134.4 | 15 022.0 |

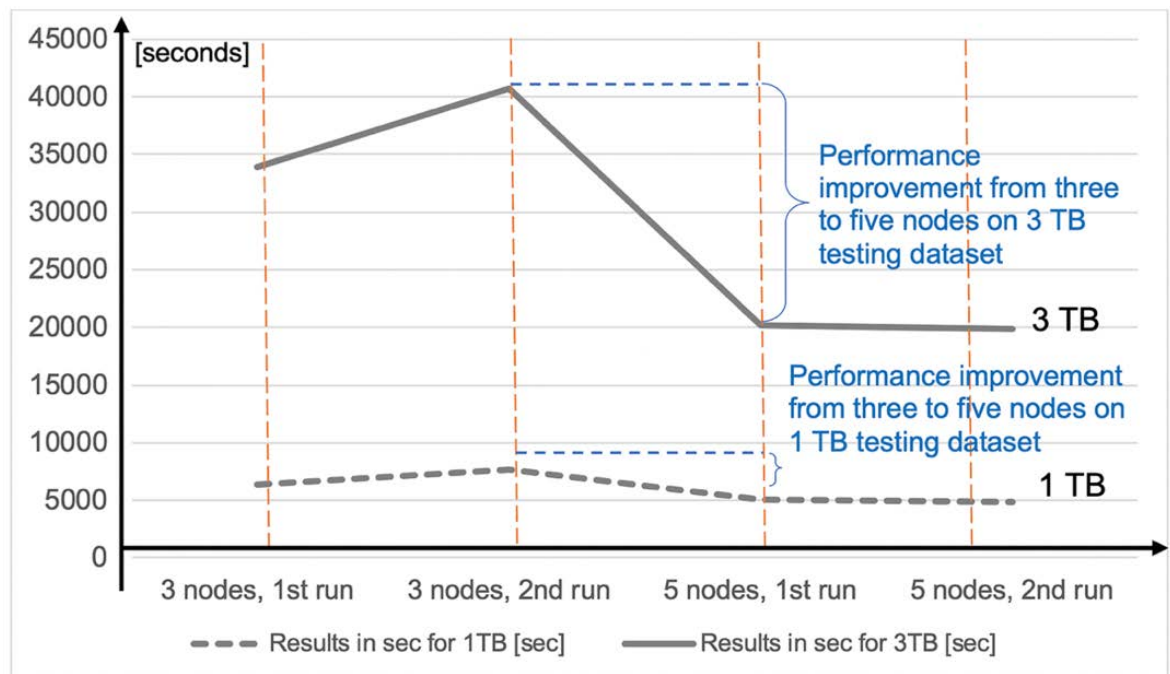

Fig. 9. Vertica cluster performance improvement for decision support queries.

From the visualisation of 1 TB and 3 TB database tests, which were matched by adding hardware resources, it is possible to conclude that for larger near-future datasets, processing demands can be matched ad-hoc by increasing hardware resources and optimising cost-effectiveness of the healthcare eSystem. For example, five nodes in the Vertica cluster (Fig. 9, Fig. 10 and Fig. 11) shows a greater cost-effective performance increase for a 3 TB (around 50%) compared to a smaller 1 TB (around 25%) database.

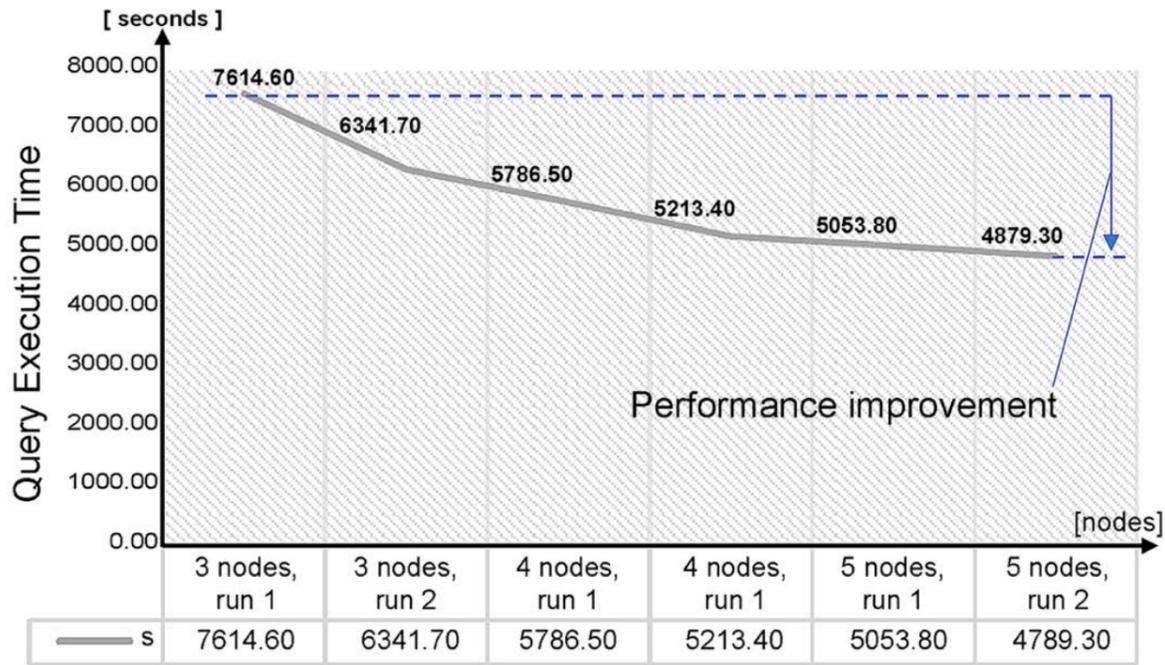

Fig. 10. Query execution time on three to five nodes in the Vertica cluster.

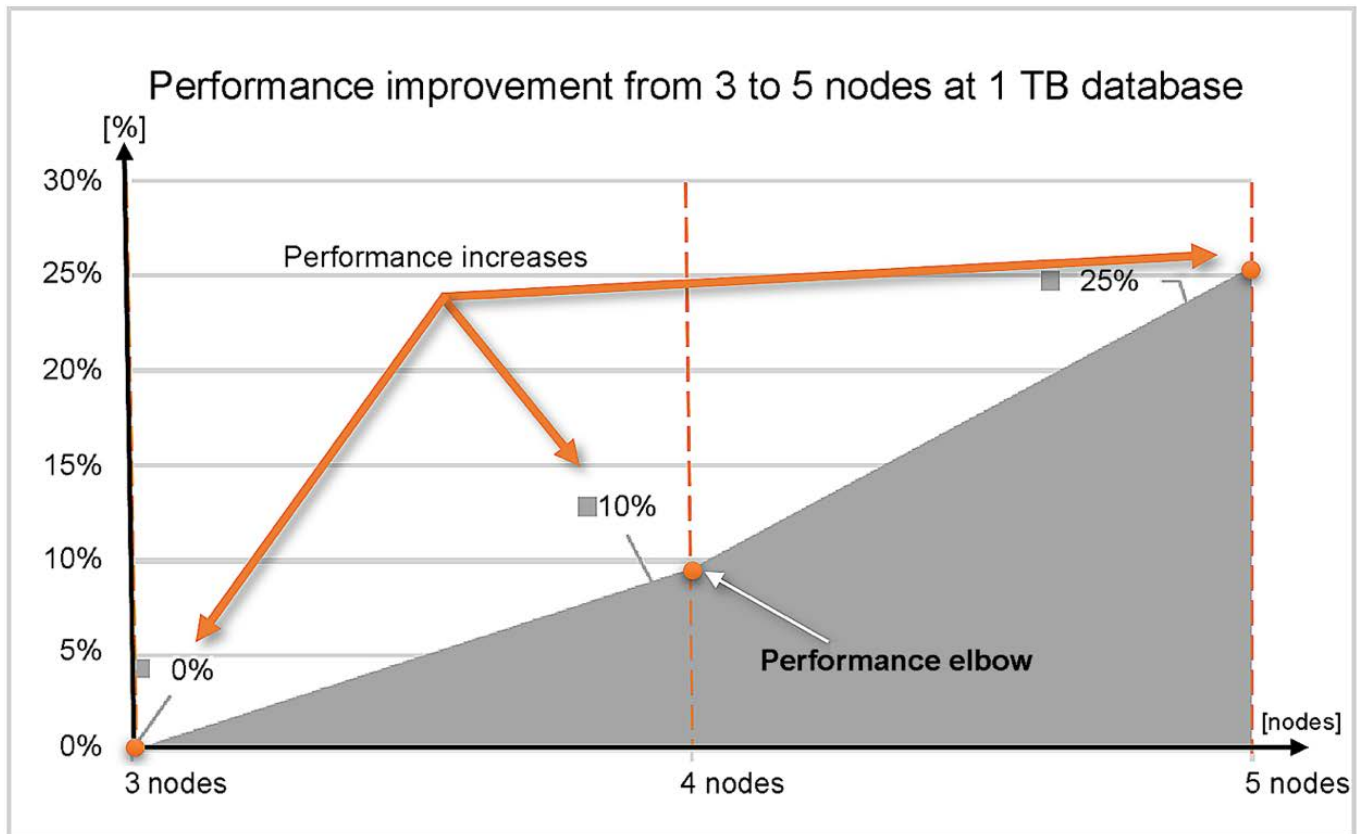

Fig. 11. Performance improvement comparing three to five nodes in the Vertica cluster for 1 TB database size.

In comparing performance improvement and scalability perspectives, the results show at least 25% performance increase from 3 to 5 nodes (Fig. 11) on 1 TB database size utilising a low-cost commodity hardware.

Query execution times and performance improvements achieved by adding extra computer resources provide sufficient evidence of a scaled-out design to work in the future with larger datasets.

Discussion

The use of big data technology intended to advance a healthcare eSystem, can be evaluated in terms of achieved performance, privacy, security, interoperability, compliance, costs and future proofing such as scalability to incremental hardware integrations, analytical tools and data increase. In the case of the Czech Republic national tender (Id. VZ0036628, No. Z2017-035520), vendor-independent solutions had to meet a large number of requirements encompassing all of the above-mentioned criteria intended to modernising the national healthcare system within the European Union. Due to contractual obligations with IHIS, as a participating party we were unable to obtain or to disseminate competitors' details including their system performance benchmarks and proposed architecture. However, our contract permits dissemination of the results and the authorship for PoC before the handover to IHIS. The presented BDA solution accepted by the Czech Republic has met all the requirements and has demonstrated system performance results well-exceeding required thresholds.

Concepts and insights transferrable to other healthcare systems are based on this case study and on the consensus of experts' views, reported literature and existing knowledge available in the public domain. The authors' views and vision for big data in future healthcare eSystems are based on professional experience, findings from Vertica-based eSystem development and big data concepts. As such, we wish to emphasise the importance of scalability for future data and performance increases, accommodation of near-future machine learning algorithms and analytical tools, security and strategic healthcare planning. Therefore, looking beyond the primary scope of this project, we question what the implications are for healthcare and other big data industry professionals. For a start, the Vertica BDA platform runs on Amazon, Azure, Google and VMware clouds providing user agility and extensibility to quickly deploy, customise, and integrate a variety of software tools. Vertica enables data warehouse transition to the cloud and to *on-premise* providing flexibility to start small and grow along with the customer's business requirements. In this case, our client (IHIS) set the conditions for implementation of the proposed solution according to the on-premise principle. The solution had to be physically isolated from the Internet and it was not possible to propose a cloud-based solution. Nevertheless, Vertica provides end-to-end security with support for industry-standard protocols, so we believe that the future of infrastructure will evolve as a multi-cloud and hybrid solution i.e. as a mixture of on-premise and cloud environments. Such data analytics and management approaches are not meant to be restricted to one type of environment only. For example, Vertica announced the availability of Eon Mode for Pure Storage (https://www.vertica.com/purestorage/) as the industry's first analytical database solution with a separation of computing and storage architecture for on-premises workload distribution.

The recommended operating system for the Vertica BDA platform is Linux Centos 7.3. Vertica also has support for other Linux-based operating systems such as (in order of authors' preference):

Red Hat Enterprise Linux (RHEL) 7.3, Oracle Enterprise Linux (OEL) 7.3, SUSE 12 SP2, Debian 8.5, and Ubuntu 14.04 LTS.

Regarding plans for our BDA solution in 2021, we are considering proposing further improvements to national healthcare and privacy protection by stream data processing from health IoT devices and mobile apps (including wearable devices such as smart watch sensors). Currently, we are conducting tests in a development environment expanded by another platform's components (Eclipse Mosquitto open source broker for carrying out stream data from IoT devices by using MQTT protocol). Acquired test data from IoT devices are transferred as stream data via MQTT Mosquitto broker (https://mosquitto.org), transformed using Apache Spark (https://spark.apache.org) and stored for future data operation purposes in Hadoop. From that layer, data are further processed in a Vertica NoSQL cluster. For IoT platform management purposes, we are using Node.js (https://nodejs.org/) to build fast and scalable network applications and the Angular platform (https://angular.io) for building mobile and desktop applications.

## Conclusion

The growing volume of medical records and data generated from near-future IoT and mobile devices mandates the adoption of big data analytics (BDA) in healthcare and related contexts. As part of the national strategy for BDA adoption in healthcare, the Czech Republic healthcare institute (IHIS) has aligned its strategy with the European Union. With over 100 complex requirements, in line with statutory regulations, included in the national public tender, was the inclusion of a subset of criteria regarding performance, cost-effectiveness, robustness and fault tolerance. Such a BDA solution had to be capable of achieving competitive and above-expected threshold results regarding overall system performance evaluation, based on TPC-H industry-standard decision support benchmark.

The tender-winning BDA solution reported here represents a snapshot in time, which exceeded expected operation on healthcare-specific TPC-H benchmark tests. The BDA solution and its control was transferred to IHIS, which over the past six month has unified the state-wide healthcare eSystem. In addition to demonstrated tests and real-life performance, the current eSystem has great potential to improve national healthcare in the Czech Republic, as well as to accommodate evolving expectations and future data needs. The produced eSystem based on Vertica analytic database management software is future proofed in terms of stream and high volume processing, scalability based on consumer/commodity hardware and fault tolerance (e.g. shutting down cluster nodes would not cause data loss). Horizontal scalability tests using commodity hardware demonstrate performance improvement of over 25% by increasing the number of cluster nodes from three to five, providing sufficient evidence of a scaled-out design based on cost-effective commodity hardware.

Currently, the produced BDA healthcare eSystem is physically isolated from the Internet infrastructure by being installed in *on premise* mode within the national geographical boundaries and therefore is considered highly secure, supporting industry standards regarding data security and protocols. The BDA healthcare eSystem supports a variety of open-source software including Linux distributions with a growing number of machine-learning libraries and integration of commercial tools such as Tableau.

The next steps in the future development of the presented healthcare eSystem includes: (1) advancement of the eSystem architecture so that the existing solution remains the blueprint architecture; (2) support for data-driven decisions during high-traffic events; (3) an increase from 100 TB to 1 PB (Petabyte) processing capability; (4) the addition of new approaches to data cleaning, storing and retrieval with minimal latency; (5) integration with other national registers (e.g. to manage and facilitate drug distribution logistics); and (6) strategic planning using healthcare data.

## Acknowledgments

We express our appreciation of the contribution of colleagues from the Solutia company who participated in the realisation of the IHIS project. Test data were obtained in collaboration with IHIS for the purposes of proof of concept development and testing. This study has been mainly supported by the Solutia company with no external financial support.

## Author Disclosure Statement

No competing financial interests exist.